\documentclass[12pt]{article}
\usepackage[dvips]{graphics}
\usepackage{amssymb}
\usepackage{amsmath}

\begin{document}


\begin{center}
{\large \bf Tau neutrino magnetic moments \\ 
from the Super-Kamiokande and $\nu$ e-scattering data} 
\end{center}
\vspace{0.5cm}

\begin{center}  
S.N.~Gninenko$^{a,b}$\footnote{e-mail address:
 Sergei.Gninenko\char 64 cern.ch} and 
N.V.~Krasnikov$^b$\footnote{nkrasnik\char 64 vxcern.cern.ch}\\
\vspace{0.5cm}
{\it $^a$CERN, Geneva, Switzerland\\
$^b$Institute for Nuclear Research of the Russian Academy of Sciences,\\ 
Moscow 117312}
\end{center}

\begin{abstract} 
  Combined results  
on $\nu_{\mu}\to \nu_{\tau}$
oscillations and  $\nu e$-scattering from the Super-Kamiokande and LAMPF
experiments, respectively,
limit the Dirac 
$\nu_{\tau}$ diagonal magnetic moment to 
$\mu_{\nu_{\tau}} < 1.9\times 10^{-9} \mu_{B}$.\  For 
the scheme with 3 Majorana neutrinos  the LAMPF results
 allow the limitation of
effective $\nu_{\tau}$ magnetic moment to $\mu_{\nu_{\tau}} < 7.6 \times 
10^{-10}\mu_{B}$.\ The moments in the scheme with additional Majorana light
 sterile neutrinos as well as experiments on
 stimulated radiative neutrino conversion are also discussed.
\end{abstract}

\section{Introduction}

In the Standard Model(SM)  neutrinos are 
massless particles with zero electric charge and magnetic moments.\ 
 Nonzero electromagnetic properties 
of neutrinos have been discussed in many extensions of the SM \cite{moh}. 
In its 
simplest extension, neutrinos acquire magnetic moments through radiative 
corrections \cite{lee,kim,marc}   
\begin{equation}
\mu_{ij} = \frac{3eG_{F}m_{ij}}{8\sqrt{2} \pi^{2}} = 3.2\times 10^{-19
}( m_{ij}/1~eV) \mu_{B} \,, 
\end{equation}  
where $G_{F}$ is the Fermi constant, $\mu_{ij} (i,j = e, \mu, \tau )$ is 
the neutrino magnetic moment matrix, $m_{ij}$ is the Dirac neutrino mass 
matrix, and $\mu_{B} = e/2m_{e}$ is the Bohr magneton. So the 
magnetic moments are too small to give any observable effect.\ However, 
there are less trivial extensions of the SM \cite{vol} predicting  
magnetic moments at the level of $(10^{-10} - 10^{-11})\mu_{B}$, that are 
large enough to be observed in neutrino electromagnetic interactions. 

 If neutrinos are Majorana
particles, their diagonal magnetic moments must be zero from the CPT 
consideration, but can have transition moments  that couple
one neutrino mass eigenstate to another.\ Dirac neutrinos can have 
both diagonal and transition moments.\ 

In the case of vacuum neutrino mixing, an evolution of the 
effective magnetic moment of the initial flavour states depends on 
whether the neutrinos are Dirac or Majorana particles, and whether 
the transition magnetic moments are involved.\
Consider the following example (see also Section 2,3).\ 
In the case of two neutrino mixing: 

\begin{eqnarray}
\nu_{\mu} = cos\theta\cdot \nu_1 + sin\theta\cdot \nu_2\\ \nonumber
\nu_{\tau} = -sin\theta\cdot\nu_1 + cos\theta\cdot\nu_2
\end{eqnarray}

where $\nu_1,\nu_2$ denote mass eigenstates and $\theta$ is a mixing angle.\
Let $\nu_1,\nu_2$ be the Dirac neutrinos with only diagonal magnetic 
moments.\ This is the scheme used by the Particle Data Group \cite{pdg}.\

The effective magnetic moments of $\mu_{\nu_{\mu}}$ and 
$\mu_{\nu_{\tau}}$ neutrino are: 

\begin{eqnarray}
\mu_{\nu_{\mu}}^2 = cos^2\theta\cdot \mu_{\nu_1}^2 + sin^2\theta\cdot 
\mu_{\nu_2}^2\\ \nonumber
\mu_{\nu_{\tau}}^2 = sin^2\theta\cdot \mu_{\nu_1}^2 + cos^2\theta\cdot 
\mu_{\nu_2}^2
\end{eqnarray}

As can be seen here,  there is no dependence of 
$\mu_{\nu_{\mu}}$ or $\mu_{\nu_{\tau}}$ on distance $L$ or neutrino 
energy $E_{\nu}$ \cite{vogel}, 
i.e. once Dirac $\nu_{\mu}$ is produced it will propagate through space
with the  value of the effective magnetic moment remaining constant,
 while the $\nu_{\tau}$ component in the beam will depend on $L$ and 
$E_{\nu}$, as described by the 
well-known formula for two neutrino mixing \cite{bil}.\  

Taking into account recent results from the Super-Kamiokande detector
on evidence for $\nu_{\mu}\to \nu_{\tau}$ oscillations with close to 
maximal mixing angle $sin^22\theta=0.8-1$ \cite{superk}, one gets

\begin{equation}
\mu_{\nu_{\mu}} \simeq \mu_{\nu_{\tau}} 
\end{equation}
   
Taking into account the upper limit to the  magnetic moment of muon neutrinos 
from the $\nu-e$ scattering experiment at LAMPF \cite{allen}:
\begin{equation}
\mu_{\nu_{\mu}} < 7.4 \times 10^{-10}\mu_B
\end{equation}
 a conservative  upper bound on the diagonal 
magnetic moment of tau neutrino can easily be derived:
\begin{equation}
\mu_{\nu_{\tau}} < 1.9 \times 10^{-9}\mu_B
\end{equation}
as well as on the magnetic moments of the mass eigenstates :
\begin{eqnarray}
\mu_{\nu_1} < 1.9 \times 10^{-9}\mu_B\\ \nonumber
\mu_{\nu_2} < 1.9 \times 10^{-9}\mu_B
\end{eqnarray}

Thus,  measurements 
of components of the neutrino mixing matrix  
would allow extraction of the fundamental magnetic moments $\mu_{ij}$.\ Note
that the limit of Eq.(6) is more then two orders of magnitude better then the 
limit extracted from 
the $\nu e$-scattering by BEBC, $\mu_{\nu_{\tau}} < 5.4 \times 10^{-7}\mu_B$
 \cite{bebc} or, limits extracted from 
the $e^+e^-$ experiments, see e.g. \cite{tristan} and its references.\\
 In experiments the neutrino originates in weak decays and is always a 
flavour eigenstate.\  Therefore, one  information on neutrino magnetic 
moments is obtained in flavour bases, while the fundamental magnetic moments 
are associated with the mass eigenstates.\
In this paper we consider the combined oscillation and magnetic 
moment effects for both  Dirac and Majorana neutrino cases  
in both mass and 
flavour eigenstate bases.\ The basic formulae are given in Section 2.\ 
In section 3 their application to 
the experimental bounds on neutrino magnetic moments and to the 
experiments on stimulated radiative neutrino conversion 
is considered.\ 
 
\section{Basic formulae}

The Lagrangian describing the neutrino interaction with the electromagnetic 
field due to non-zero anomalous transition magnetic and dipole moments 
in the mass eigenstates basis takes the form 
\begin{equation}
L_{\mu} = \frac{1}{2} \sum_{i,j = 1,N} \bar{\nu}_j \sigma_{\mu \nu}
(\mu_{ji} +d_{ji}\gamma_{5}) \nu_{i} F_{\mu \nu} +h.c.
\end{equation}
Here  $\nu_{i}$ and $\nu_{j}$ are neutrinos with masses $m_i$ and 
$m_j$ respectively.\ The Lagrangian (8) can be rewritten in the form
\begin{equation}
L_{\mu} = \frac{1}{2}\sum_{i,j = 1,N}\bar{\nu}_{jR}{\mu}_{jiRL}
\sigma_{\mu\nu}\nu_{iL}F_{\mu\nu} + h.c.
\equiv \frac{1}{2}\bar{\nu}_R\hat{\mu}_{RL}
\sigma_{\mu\nu}\nu_LF_{\mu\nu} +h.c. \,,
\end{equation} 
where $\nu_{iL,R} = \frac{1}{2}(1 \mp \gamma_5)\nu_i$ and 
\begin{equation}
\mu_{jiRL} = (\mu -d)_{ji} + (\mu +d)^{*}_{ij}
\end{equation}

For the Dirac neutrino (for the Majorana neutrino we have to substitute
$\mu_{ji} \rightarrow 2 Im(\mu_{ji})$, $d_{ji} \rightarrow 
2Im(d_{ji})$)  
the decay width $\nu_{i} \rightarrow \nu_{j} + \gamma$ is 
determined by the formula 
\begin{equation}
\Gamma(\nu_{i} \rightarrow \nu_{j} + \gamma) = 
\frac{m_{i}^{3}(| \mu_{ji}|^2 +|d_{ji}|^2)}  {8\pi}
(1 -\frac{m_j^2}{m_i^2})^3~~,
\end{equation}
Or in terms of inverse radiative lifetime formula (11) takes the form
\begin{equation}
\tau^{-1}_{\gamma} = 5.3s^{-1}(\frac{\mu_{ji,eff}}{\mu_B})^2 
(\frac{m^2_i -m^2_j}{m^2_i})^3(\frac{m_i}{eV})^3, 
\end{equation}
where $\mu^2_{ji,eff} = |\mu^2_{ji}| + |d^2_{ji}|$.
In general, there are other electromagnetic form-factors besides 
magnetic and dipole moments but due to gauge invariance only 
magnetic and dipole form-factors contribute to radiative decays \cite{moh}.
Nonzero magnetic and dipole neutrino  moments lead in  
particular to additional contributions to  weak 
neutrino-electron scattering, namely for $E_e \gg m_i,m_j$ 
\begin{equation}
\frac{d\sigma}{dE_e} = \frac{d\sigma_{st}}{dE_e} + 
\frac{d\sigma_{\mu}}{dE_e}\,,
\end{equation}
where $\frac{d\sigma_{st}}{dE_e}$ is the standard  weak 
scattering cross section and 
\begin{equation}
\frac{d\sigma_{\mu}}{dE_e} = \frac{\pi\alpha^2}{m^2_e}
\frac{(|\mu_{\nu, eff}|^2 }{\mu_B^2}(\frac{1}{E_e- m_e} - 
\frac{1}{E_{\nu}})
\end{equation} 
is the  additional contribution to the weak cross section  
from nonzero magnetic and dipole neutrino moments. In formula (14)  
$\mu_{\nu,eff}$  is determined by nonzero magnetic and 
dipole moments.\
These two contributions are incoherent within the limits of vanishing neutrino 
mass  since weak scattering 
preserves the neutrino helicity whereas magnetic 
scattering changes it helicity.\ Eqs.(8,9) are
 written  in the neutrino eigenstates mass basis. 

 Now, consider the case of Majorana 
neutrinos.\  In the Weyl basis 
the interaction (9) can be rewritten in the form
\begin{equation}
L_{\mu} = \frac{1}{4} \sum_{i,j =1,N}\bar{\nu}^c_{L,i}\sigma_{\mu\nu} 
\tilde{\mu}_{ij}^M \nu_{L,j}F_{\mu\nu} +h.c.
\end{equation}
The magnetic moment matrix $\tilde{\mu}_{ij}^M$ is antisymmetic 
$\tilde{\mu}_{ij}^M = -\tilde{\mu}_{ji}^M$ 
for Majorana neutrinos \cite{moh} and 
$\nu^c_L \equiv (\nu_L)^c = \frac{1 + \gamma_5}{2}\nu^{c}$, 
$\nu^c = C\gamma^0\nu^{*}$, $C = i\gamma^{2}\gamma^{0}$, 
$\bar{\nu}^c = \nu C$.
Here $\nu_{i}$ are the neutrinos in the eigenstates mass basis with masses 
$m_i$. The realistic case corresponds to N=3 
(electron, muon and tau neutrinos). 
However, recent results from a solar neutrino search, Super-Kamiokande and 
LSND results favour the  very intriguing possibilty of the 
existence of a 4-th light neutrino (sterile neutrino) \cite{valle}, so maybe 
$N >3$.

For the scattering of the neutrino $\nu_i$ with definite mass on the target, 
the contribution to the total cross section from a nonzero neutrino magnetic 
moment matrix is 
\begin{equation}
\sigma_{\mu}(\nu_i e \rightarrow ...) = \sum_{j}\sigma_{\mu}
(\nu_i e \rightarrow \nu_j e) \sim \sum_j |\mu^M_{ji}|^2 = 
((\hat{\mu}^M)^+\hat{\mu}^M)\,,
\end{equation}
where $\hat{\mu}^M_{ij} = \tilde{\mu}^M_{ij}$.  

Both here and elsewhere in this paper
 we ignore spins and phase space factors.\ 
 The total cross section is
the sum of different mass eigenstate transitions combined incoherently,
since in principle their contributions are distinguishable in the
 final state.\   

The relation among neutrino states  in flavour and 
mass eigenstates bases is 
determined by the neutrino mixing matrix, namely \cite{bil}
\begin{equation}
|\nu^{F}_{iL}> = \sum_{j =1,N}U_{ij}|\nu_{jL}>
\end{equation}
Here $|\nu^{F}_{iL}>$ are neutrino states in flavour basis ($|\nu^{F}_{1L}> 
=|\nu_{eL}>$, $|\nu^F_{2L}> = |\nu_{\mu L}>$, $|\nu^F_{3L}> = |\nu_{\tau L}>$, 
...). Note that we have written the expression (17) at $t = 0$, at 
arbitrary time $t$ the matrix $U_{ij}$ depends on time or 
on the distance $l$ between the source of neutrinos and detector  
(neutrino oscillations phenomenon), namely \cite{bil} 
\begin{equation}
U_{ij}(l) = U_{ij}\exp(-iE_{j}l) \approx \exp(-iEl) \times 
U_{ij}\exp [-i(m^2_j/2E)l]  
\end{equation}
Here $U_{ij}$ and $U_{ij}(l)$ are unitary matrices.

For the scattering of the neutrino $\nu^F_i$ with definite flavour 
according to the principle of superposition, the 
amplitude $A(\nu_i^F e \rightarrow \nu_j e)$ is proportional to 
\begin{equation}
A(\nu^F_i e \rightarrow \nu_j e) \sim 
\sum_{k}\mu^M_{jk}U_{ik}
\end{equation}
The total cross section is proportional to 
\begin{equation}
\sigma_{\mu}(\nu_i^Fe \rightarrow...) = 
\sum_j\sigma_{\mu}(\nu_i^Fe \rightarrow \nu_je) 
\sim \sum_j|\sum_k \mu^M_{jk}U_{ik}|^2 \equiv |\mu^F_{eff,i}|^2
\end{equation}
One can rewrite $|\mu^F_{eff,i}|^2$ in the form 
\begin{equation}
|\mu^F_{eff,i}|^2 = (\hat{U}^{T+}\hat{\mu}^{M+}
\hat{\mu}^M \hat{U}^T)_{ii} = 
((\hat{\mu}^{F})^{+}\hat{\mu}^F)_{ii} ,\,
\end{equation}
where 
\begin{equation}
\hat{\mu}^F = (\hat{U^T})^{+}\hat{\mu}^M\hat{U}^T \,,
\end{equation} 
$(\hat{U})_{ij} = U_{ij}$, $(\hat{U}^T)_{ij} = (\hat{U})_{ji}$, 
$(\hat{U}^+)_{ij} = (\hat{U})_{ji}^{*}$,   
$(\hat{\mu}^M)_{ij} = \tilde{\mu}^M_{ij}$
 
Experimentally, neutrinos are  produced  in 
flavour states (e.g., $\nu_{eL} $ or $\nu_{\mu L}$ neutrinos). 
If the distance between 
the source and detectror is not small they will oscillate into 
other flavour states. In matrix notations we have the following 
equation relating 
neutrino wave function at $l =0$ and at  $l = ct$

\begin{equation}
|\nu^F_{Li}(l)> = \sum_{j}(\hat{U}^{'}(l))_{ij}|\nu^F_{Lj}(l=0)> \,,
\end{equation}   

where
$$
\hat{U}^{'}(l) = \hat{U}(l)\hat{U}^{-1}(l=0)
$$
For the scattering of the neutrino state $|\nu^F_{Li}(l)>$, which is 
a mixture of the flavour neutrino states, the formulae (21-22) 
take place with the substitution of $\hat{U} \rightarrow \hat{U}(l)$.

For instance, formula (20) takes the form 

\begin{eqnarray}
\sigma_{\mu}(\nu^F_{Li}(l)\,e \rightarrow...) = \sum_{j}
\sigma_{\mu}(\nu^F_{Li}(l)\,e \rightarrow \nu_j\, e) \sim
\sum_j |\mu^F_{ji}(l)|^2\\ \nonumber
 = ((\hat{\mu}^F(l))^+\hat{\mu}^F(l))_{ii} \equiv |\mu^F_{eff,i}(l)|^2,
\end{eqnarray}

where 

\begin{equation}
\hat{\mu}^F(l) = (\hat{U}(l)^T)^+\hat{\mu}^M\hat{U}(l)^T \,,
\end{equation}
\begin{equation}
\hat{\mu}^F(l=0) = \hat{\mu}^F
\end{equation} 
 
Thus, for example, in the Majorana case, the total cross section for the magnetic 
scattering can be interpreted as 
a  sum of cross sections corresponding to the {\it initial} to
{\it final} flavor transitions combined incoherently,
since in principle their contributions are distinguishable in the
 final state.\   
In general, there will be dependence of the effective magnetic moment 
on the distance l in flavour bases \footnote{This fact is also mentioned
in ref.\cite{vogel}}.
 
For the case $N = 2$, the neutrino magnetic  mass matrix 
has the form $\tilde{\mu}^F_{ij} = \epsilon_{ij} \mu$ ($\epsilon_{12} = 
- \epsilon_{21} =1 , \epsilon_{11} = \epsilon_{22} = 0$) and 
$\mu^{F}\mu^{F+} = |\mu^2|\times I$, where $I$ is unit matrix. 
Therefore  the magnetic transition moment $|\mu|$ in 
flavour and mass eigenstates 
bases coincides and   
does not depend on  distance $l$. For the most realistic case $N =3$ 
neutrino magnetic matrix $\tilde{\mu}^F_{ij}$ has three nonzero matrix 
elements 
$\tilde{\mu}^F_{12} \equiv \mu_{e\mu}$, $\tilde{\mu}^F_{13} 
\equiv \mu_{e\tau}$ and 
$\tilde{\mu}^F_{23} \equiv \mu_{\mu \tau}$ and for $\nu_e-$, 
$\nu_{\mu}-$, $\nu_{\tau}-$ electron 
scattering in accordance with formulae (24-26), where  
the corresponding effective  moments are
\begin{equation}
\mu^2_{\nu_e,eff} = |{\mu}_{e\mu}|^2 + |{\mu}_{e\tau}|^2,
\end{equation}
\begin{equation}
\mu^2_{\nu_{\mu},eff} = |{\mu}_{e\mu}|^2 + |{\mu}_{\mu\tau}|^2,
\end{equation}
\begin{equation}
\mu^2_{\nu_{\tau},eff} = |{\mu}_{e\tau}|^2 + |{\mu}_{\mu \tau}|^2
\end{equation}

Trivial inequality takes place
\begin{equation}
\mu^2_{\nu_{\tau},eff} \leq \mu^2_{\nu_e,eff} + \mu^2_{\nu_{\mu},eff}
\end{equation}

The trace of $\hat{\mu}^{F+}\hat{\mu}^{F}$   
is invariant under unitary transformations linking flavour  and mass 
eigenstates bases.\  
For the pattern involving 3 Majorana neutrinos, therefore the 
following combination 
does not depend on the distance l and coincides for eigenstates mass  
and flavour bases:
\begin{equation}
I_3 \equiv \frac{1}{2} Tr(\hat{\mu}^{F+}\hat{\mu}^{F}) = 
 |\mu_{e\mu}|^2  + |\mu_{e\tau}|^2 + |\mu_{\mu\tau}|^2
\end{equation}

Note that the schemes with an additional light sterile neutrino 
\cite{pel} are now rather popular due to the fact that  3 neutrinos scheme 
cannot explain simultaneously atmospheric, solar and LSND 
results. In the scheme with Majorana 
neutrinos it corresponds to the case $N >3$. Consider the case $N = 4$ 
(with an additional light sterile neutrino). For such a case, in addition to 
nonzero matrix elements for 
$3 \times 3$ matrix we have nonzero matrix elements $\mu_{es}$, $\mu_{\mu s}$, 
$\mu_{\tau s}$ describing transitions of $\nu_e$, $\nu_{\mu}$, $\nu_{\tau}$ 
neutrinos to the sterile neutrino $\nu_s$, and in the 
formulae (27-29) we have to 
add $|\mu^2_{es}|$, $|\mu^2_{\mu s}|$ and $|\mu^2_{\tau s}|$ on the 
right-hand sides of the equations (27), (28) and (29), respectively.\ 
The corresponding invariant (analogue of (31)) is 
\begin{equation}
I_4 = I_3 + |\mu_{es}|^2 + |\mu_{\mu s}|^2 + |\mu_{\tau s}|^2
\end{equation}

Consider briefly the case of Dirac neutrinos. The derivation of the 
main formulae 
is similar to the Majorana neutrino case except for two differences. 
The first is  that the 
neutrino magnetic matrix $\hat{\mu}_{RL}$ in formula (9) 
is not asymmetric and in general 
is arbitrary.\ The second difference is that diagonalization of 
neutrino mass matrix is performed as an independent unitary rotation 
of left-handed and right-handed neutrinos by two 
independent unitary matrices $\hat{U}_{L}$ and $\hat{U}_{R}$. 
The relation among 
neutrino magnetic moments in mass eigenstates and in flavour bases 
has the form 
\begin{equation}
\hat{\mu}^F_{RL} = \hat{U}^{T+}_R \hat{\mu}_{RL}\hat{U}_{L}^{T}
\end{equation}
Note that in general  the flavour basis for the 
right-handed neutrino is not well 
defined and can be arbitrary. So we can work in a basis where 
$\hat{U}_R = 1$, but the main 
formulae for effective magnetic moments in the flavour basis do not depend 
on $\hat{U}_R$. The effective magnetic moments for flavour neutrinos are 
determined by formulae analogous to (20-26), namely
\begin{equation}
\mu^2_{{\nu}_i,eff} = (\hat{\mu}^{F+}_{RL}\hat{\mu}^{F}_{RL})_{ii} = 
(\hat{U}^{T+}_L\hat{\mu}^{+}_{RL}\hat{\mu}_{RL}\hat{U}_T)_{ii} 
\end{equation}
For instance, if in massive eigenstates basis Dirac 
neutrinos are diagonal (this 
scenario is used by the Particle Data group \cite{pdg}), i.e. 
\begin{equation}
\mu_{ij,RL} = \mu_j \delta_{ij},
\end{equation}
then in the flavour basis we have
\begin{equation}
\mu^2_{{\nu}_i,eff} = \sum_{j}|U_{ij}|^2|\mu_j|^2
\end{equation}
and there is no dependence on the distance l or neutrino energy for effective 
magnetic moments (36) \cite{vogel}.    
In the general case , the 
effective magnetic moments for  $\nu_e$, $\nu_{\mu}$ and 
$\nu_{\tau}$ electon scattering are determined by the 
flavour magnetic moment matrix, namely  
 
\begin{equation}
\mu^2_{\nu_e, eff} = |\mu^F_{ee,RL}|^2 + |\mu^F_{\mu e, RL}|^2 + 
|\mu^F_{\tau e , RL}|^2,
\end{equation}
\begin{equation}
\mu^2_{\nu_{\mu},eff} = |\mu^F_{\mu \mu ,RL}|^2 + |\mu^F_{e \mu  , RL}|^2 + 
|\mu^F_{\tau \mu , RL}|^2,
\end{equation}
\begin{equation}
\mu^2_{\nu_{\tau}, eff} = |\mu^F_{\tau \tau, RL}|^2 + 
|\mu^F_{e \tau , RL}|^2 + |\mu^F_{\mu \tau  , RL}|^2
\end{equation}
Note that in general $\mu_{e \mu , RL} \neq \mu_{\mu e, RL}$. 
For the case of Dirac neutrinos the corresponding 
invariant (31) is 
\begin{equation}
Tr(\hat{\mu}^{F+}\hat{\mu}^{F}) = \mu^2_{\nu_e,eff} + \mu^2_{\nu_{\mu},eff} 
+\mu^2_{\nu_{\tau},eff}
\end{equation}

\section{Experimental bounds on magnetic moments}

In this section we consider the following bounds on neutrino magnetic 
moment matrix 
elements resulting from reactor and accelerator experiments:
\footnote{The astrophysical bounds on neutrino magnetic moments can 
be found in ref.\cite{raffelt}.}
 
\begin{equation}
\mu_{\nu_{e}} < 1.8 \times10^{-10}\mu_B
\end{equation} 

\begin{equation}
\mu^2_{\nu_{e}} + 2.1\mu^2_{\nu_{\mu}} < 1.16 \times 10^{-18} \mu^2_B
\end{equation}

\begin{equation}
\mu_{\nu_{\tau}} < 5.4 \times 10^{-7}\mu_B
\end{equation}

from the reactor \cite{derbin}, $\nu e$ scattering at LAMPF \cite{allen} and
BEBC \cite{bebc} data, respectively.\ The analysis 
of Super-Kamiokande results allows to obtain similar or more 
stringent bounds \cite{vogel}, \cite{pulido}, \cite{sng}:
\footnote{It should be noted that 
the bound of ref.\cite{sng} is obtained in the flavour bases.}
 
\begin{eqnarray}
\mu_{\nu_{e}} < (1.6-2) \times10^{-10}\mu_B\\ \nonumber
\mu_{\nu_{\tau}} < 1.6 \times 10^{-7} \mu_B
\end{eqnarray}

In the derivation of the bounds (41-43) the distance between the source 
and the detector was small compared to possible neutrino oscillation length 
so the possible influence of the neutrino oscillation on the extraction of 
bounds on neutrino magnetic moments
(the dependence of the extracted magnetic moments 
on the distance between neutrino source and detector) 
is  negligible. In experiments we measure inclusive cross sections of 
the flavour neutrinos scattering on target, so the experimental 
bounds  (41-43) on diagonal neutrino magnetic moments are also valid for 
the corresponding effective flavour magnetic moments which are 
generalizations of the diagonal magnetic moments for the case of nondiagonal 
magnetic moments.    

Consider at first the case of the 
3 Majorana neutrinos.\ Using reactor bound (41) and formulae of Eqs.(27,28)
we find that 
\begin{equation}
|\mu_{e\mu}|,|\mu_{e\tau}| < 1.8\times 10^{-10}\mu_B
\end{equation}
LAMPF bound (42) leads to 
\begin{equation}
|\mu_{e\mu}| < 6.1 \times 10^{-10}\mu_B \,,
\end{equation}
\begin{equation}
|\mu_{e\tau}| < 1.1\times 10^{-9}\mu_B \,,
\end{equation}
\begin{equation}
|\mu_{\mu\tau}| < 7.4 \times 10^{-10}\mu_B \,,
\end{equation}

Bounds (45-48) have been obtained  for magnetic moments in the 
flavour bases.
Note that from the inequalities (45-48) we find that

\begin{equation} 
I_3 = |\mu_{e\mu}|^2 + |\mu_{e\tau}|^2 + |\mu_{\mu \tau}|^2 < 
0.61 \times 10^{-18}\mu^2_B
\end{equation}

As shown in section (2) this invariant is 
independent on distance l between the neutrino source 
and detector and is the same for mass eigenstates 
and flavour eigenstates bases. 
So we find that in any basis, nondiagonal neutrino transition moments 
in scheme with 3 Majorana neutrinos have to be less than

\begin{equation}
|\mu_{ij}| < 7.8 \times 10^{-10}\mu_{B}
\end{equation}

In particular, from the inequality (49) and formula of Eq.(29) we find that 
in a scheme with 3 Majorana neutrinos  
the effective magnetic moment of $\tau-$neutrino has to be less than 

\begin{equation}
\mu_{\nu_{\tau},eff} < 7.6 \times 10^{-10}\mu_B
\end{equation} 

For the scheme with additional sterile Majorana neutrinos, bounds 
(45-48) are also valid.\ From the reactor bound (41) we find that 
$|\mu_{es}| < 1.8 \times 10^{-10}\mu_B$. From the LAMPF bound Eq.(42) 
we find that  $|\mu_{es}| < 1.1 \times 10^{-9}\mu_B$ and 
$|\mu_{\mu s}| < 7.4 \times 10^{-10}\mu_B$. From BEBC bound (43) we find 
that $|\mu_{\tau s}| < 5.4 \times 10^{-7}\mu_B$. Using the formula (32) 
one can find that in any basis (mass eigenvalue basis 
for instance) nonzero transition magnetic moments in the
scheme with  additional light Majorana neutrinos have to be less than

\begin{equation}
|\mu_{ij}| < 5.4 \times 10^{-7}\mu_B
\end{equation}

For 3 Dirac neutrinos, in addition to 
bounds on diagonal magnetic moments 
similar bounds on the transition magnetic moments can be deduced. 
Namely, in all bases, 
nondiagonal neutrino magnetic moments have to be less than $5.4 \times
10^{-7}\mu_B$.  It should be noted that bound (44) on a diagonal 
magnetic moment of Dirac $\tau$-neutrino obtained from 
Super-Kamiokande data in the assumption of the nearly maximal $\nu_{\tau} - 
\nu_{\mu}$ mixing leads to the same bound on  
$|\mu_{\tau s}|$ in the model with 4 Majorana neutrino 
($\nu_e$, $\nu_{\mu}$, $\nu_{\tau}$ and $\nu_{s}$). For the model with 
3 Dirac neutrino the bound of (44) gives a similar bound on 
transition magnetic 
moments $\mu_{e \tau }$, $\mu_{\mu \tau}$. For the interpretation of 
Super-Kamiokande data as $\nu_{\mu} - \nu_{s}$ oscillation using the results 
of ref.\cite{bebc} it can be found that in 4-Majorana neutrino model bound 
(47) also takes place.\ Note, that bounds on heavy sterile neutrino 
transition magnetic moments were obtained  from the NOMAD data in 
ref. \cite{gnin}.\ 

For 3 Dirac neutrinos in the assumption of the diagonal magnetic moments in the
mass eigenstates basis and with maximal $\nu_{\mu} -\nu_{\tau}$ mixing 
\footnote{The assumption of nearely maximal $\nu_{\mu} -\nu_{\tau}$ mixing 
is the simplest interpretation of Super-Kamiokande data} we find that 
$|U_{1j}|^2 \approx \delta_{1j}$, $|U_{22}|^2 \approx |U_{23}|^2 
\approx |U_{32}|^2 \approx |U_{33}|^2 \approx 0.5$, $|U_{21}|^2 \approx 
|U_{31}|^2 \approx 0$ and as a consequence 
\begin{equation}
\mu^2_{\nu_{\tau},eff} \approx  \mu^2_{\nu_{\mu},eff} 
\approx \frac{1}{2}(\mu^2_2 + 
\mu^2_3)
\end{equation}
Using the limit (42) we obtain the bounds of Eq.(6,7) discussed above.

There are  interesting proposals  on the
search for radiative neutrino transitions through 
neutrino flavour conversion in a superconducting 
RF cavity installed in a neutrino beam \cite{mat}-\cite{vann2}.\
Consider now the application of the obtained bounds on the neutrino 
radiative decays taking the example of  the model with 3 Majorana neutrinos. 
It follows from Eqs.(12,50) that 

\begin{equation}
\tau(\nu_i \rightarrow \nu_j + \gamma) > 3 \times 10^{17}s \times 
(\frac{m^2_i}{|m^2_j - m^2_i|})^3(\frac{eV}{m_i})^3 
\end{equation}

If neutrino masess $m_i$ and $m_j$ are degenerate,  
suppression factors for the neutrino radiative lifetime arises dramatically.\
For example, for $(\frac{|m^2_j - m^2_i|}{m^2_i})\simeq 10^{-3}$ and 
$m^2_i \simeq 1~eV^2$ 
the radiative neutrino lifetime is 
$\tau(\nu_i \rightarrow \nu_j + \gamma)\gtrsim 10^{26}$ sec.\ 
This estimate, 
compared  with expected sensitivity of $\lesssim 10^{20}$ sec for the 
same mass values in the proposed experiments,
 makes  quite difficult for them  to compete with the bound of Eq.(50).\   
\vspace{1.0cm}

{\large \bf Acknowledgements}\\

We would like to thank Prof. V.A. Matveev  for the interesting 
discussion he had with us.\ 
The work of N.V.K. has been partly supported in Russia 
by RFFI grants 99-02-16956 and 99-01-00091.


\vspace{1.0cm}



\begin{thebibliography}{99}

\bibitem{moh} For a review see R.N. Mohapatra and  P.B. Pal, in: "Massive 
Neutrino in Physics and Astrophysics", World Scientific, Singapore, 1991.

\bibitem{lee} B.W. Lee and R.E. Shrok, Phys. Rev. {\bf D16} (1977) 1444.

\bibitem{kim} J. E. Kim, Phys. Rev. Lett. {\bf 41} (1978) 360; 
preprint hep-ph/9904312.

\bibitem{marc} W. Marciano and A.I. Sanda, Phys. Lett. {\bf B67} (1977) 303.

\bibitem{vol} M.B. Voloshin, Sov. J. Nucl. Phys. {\bf 48} (1988) 512; \\
R. Barbieri and R.N. Mohapatra, Phys. Lett. {\bf B218} (1989) 225; \\
K.S. Babu and R.N. Mohapatra, Phys. Rev. Lett. {\bf 63} (1989) 228.

\bibitem{pdg} Review of Particle Physics, The European Physical Journal 
C3 (1998) 1.


\bibitem{vogel} J.F. Beacom and P. Vogel,  
Phys. Rev. Lett.{\bf 83} (1999) 5222. 

\bibitem{bil} See e.g. review: S.M. Bilenky and B. Pontekorvo, Phys. Rep. {\bf 41}
(1978) 225. 





\bibitem{superk}
Super-Kamiokande Collab., Y.~Fukuda et al., Phys. Lett. B433 (1998) 9; \\
Phys. Rev. Lett. {\bf 81} (1998) 1562; Phys. Lett. {\bf B436} (1998) 33.



\bibitem{allen} R.C. Allen et al., Phys. Rev. {\bf D47} (1993) 11.

\bibitem{bebc} A.M. Cooper-Sarkar et.al., Phys. Lett. {\bf B280} (1992) 153.

\bibitem{tristan} N. Tanimoto, I. Nakano and M. Sakuda, 
preprint hep-ph/0002170.

\bibitem{pel} J.T. Peltoniemi and J.W.F. Valle, Nucl. Phys. {\bf B406} (1993) 409; \\
D.O. Caldwll and R.N. Mohapatra, Phys. Rev. {\bf D48} (1993) 3259. 

\bibitem{raffelt} See for instance: G.G. Raffelt, in Proceedings 1998 Summer 
School in High-Energy Physics and Cosmology, ICTP,Trieste,Italy \\ 
(ed. by G. Senjanovic and A.Yu. Smirnov, World Scientific, Singapore). 

\bibitem{derbin} A.I. Derbin et al., JETP Lett., {\bf 57} (1993) 768.


\bibitem{sng} S.N. Gninenko, Phys. Lett. {\bf B452} (1999) 414.


\bibitem{pulido} J. Pulido and A. M. Mourao, Nucl. Phys. B(Proc. Suppl.)
 {\bf 70} (1999) 255.

\bibitem{kang} Sin Kyu Kang, Jihn E. Kim and Jae Sik Lee, Phys.Rev. {\bf D60}
(1999) 033008.

\bibitem{valle} As a recent review see, e.g., Jose W.E. Valle, hep-ph/9911224.

\bibitem{gnin} S.N. Gninenko and N.V. Krasnikov, Phys. Lett. {\bf B450} (1999) 165.

\bibitem{mat} S. Matsuki and K. Yamamoto, Stimulated conversion \\
of neutrinos: a new method to search for radiative decays \\
ICRNP-Kyoto University PREPRINT-9203(1992).

\bibitem{vann1} M.C. Gonsalez-Garcia, F. Vannucci and J. Castromonte,  \\
Phys. Lett. B373 (1996) 153.

\bibitem{vann2} F. Vannucci, preprint hep-ex/9911025. 

\end{thebibliography}
\end{document}